\documentstyle[12pt,epsf]{article}
\newcommand{\beq}{\begin{equation}}
\newcommand{\eeq}{\end{equation}}
\newcommand{\beqa}{\begin{eqnarray}}
\newcommand{\eeqa}{\end{eqnarray}}
\newcommand{\ba}{\begin{array}}
\newcommand{\ea}{\end{array}}
\newcommand{\CR}{\nonumber \\}

\newcommand{\A}{\alpha}
\newcommand{\B}{\beta}
\newcommand{\D}{\delta}

\newcommand{\La}{\Lambda}

\newcommand{\lm}{\lambda}

\newcommand{\tQ}{\tilde{Q}}

\newcommand{\la}{{\langle}}
\newcommand{\ra}{{\rangle}}

\begin{document}

\begin{titlepage}
\null
\begin{flushright} 
hep-th/9808022  \\
UTHEP-385  \\
August, 1998
\end{flushright}
\vspace{0.5cm} 
\begin{center}
{\Large \bf
Seiberg-Witten Geometry with \\ Various Matter Contents
\par}
\lineskip .75em
\vskip2.5cm
\normalsize
{\large Seiji Terashima and Sung-Kil Yang} 
\vskip 1.5em
{\large \it Institute of Physics, University of Tsukuba \\
Ibaraki 305-8571, Japan}
\vskip3cm
{\bf Abstract}
\end{center} \par
We obtain the Seiberg-Witten geometry for four-dimensional $N=2$ gauge theory
with gauge group $SO(2N_c)$ $(N_c \leq 5)$ with massive spinor and vector 
hypermultiplets by considering the gauge symmetry breaking in the $N=2$
$E_6$ theory with massive fundamental hypermultiplets. 
In a similar way the Seiberg-Witten geometry is determined for $N=2$ $SU(N_c)$
$(N_c \leq 6)$ gauge theory with massive antisymmetric and fundamental 
hypermultiplets. Whenever possible we compare our results expressed in the 
form of ALE fibrations with those obtained by geometric engineering and 
brane dynamics, and find a remarkable agreement. We also show that these 
results are reproduced by using $N=1$ confining phase superpotentials.

\end{titlepage}

\baselineskip=0.7cm

%%%%%%%%%%%%%%%%%%%%%%%%%%%%%%%%%%%%%%%%%%%%%%%%%%%%%%%%%%%%%%%%%%%%%%
\section{Introduction}
%%%%%%%%%%%%%%%%%%%%%%%%%%%%%%%%%%%%%%%%%%%%%%%%%%%%%%%%%%%%%%%%%%%%%%

In the seminal paper by Seiberg and Witten (SW),
it was discovered that the low-energy behavior of $N=2$ $SU(2)$
supersymmetric gauge theory in four dimensions is described in terms of the
geometry associated with the Riemann surface \cite{SeWi}. Extensions of their
work to the case of other gauge groups were carried out by several 
groups \cite{ArFa}-\cite{Ha}. At first sight it was unclear 
if the Riemann surface in the exact description is an auxiliary object
for mathematical setup or a real physical object.
It turns out that four-dimensional $N=2$ gauge theory on ${\bf R}^4$ 
is realized by an M-theory fivebrane on
${\bf R}^4 \times \Sigma$ embedded in ${\bf R}^{10} \times S^1$,
where $\Sigma$ is the SW Riemann surface \cite{KlLeMaVaWa,Wi}.

Gauge theories associated with the configuration ${\bf R}^4 \times \Sigma$
can be analyzed by the brane dynamics \cite{GiKu}.
So far $N=2$ gauge theories with classical gauge groups 
containing flavor matters in the fundamental representation
have been understood successfully in this framework \cite{LaLoLo1}.
Toward further developments
it is highly desirable to be able to describe 
matter contents in various representations as well as 
exceptional gauge symmetry.
However, it is still difficult to explain the exceptional 
gauge symmetry along the lines of \cite{Wi}.
Concerning the matter representations other than the fundamentals,
our analysis has so far been restricted to the case of 
$N=2$ $SU(N_c)$ gauge theory with matters in the symmetric or antisymmetric 
representations \cite{LaLo,LaLoLo2}.
The difficulty lies not only due to the lack of precise knowledge of the 
brane dynamics, but also due to the lack of the field theory answer.

Our purpose in this paper is to present some field theory answer to 
the above issue.
Staring with the $N=2$ SW geometry with $E_6$ gauge group with 
massive fundamental matters, 
which we proposed in the previous communication \cite{TeYa3},
we construct the SW geometry with $SU(N_c)$ and $SO(N_c)$ gauge groups
with various matter contents. All these geometries we will obtain take
the form of a fibration of the ALE spaces over a sphere.

This paper is organized as follows.
In sect.2, we explain in detail how to implement the gauge symmetry breaking 
in the SW geometry by giving appropriate VEV to the adjoint scalar field in
the $N=2$ vector multiplet.

In sect.3, breaking the $E_6$ symmetry down to $SO(2 N_c)$ $(N_c \leq 10)$,
we derive the SW geometry for $N=2$ $SO(2 N_c)$ theory with massive 
spinor and vector hypermultiplets.
In the massless limit, our $SO(10)$ result is in complete agreement with the
one obtained by the method of geometric engineering \cite{AgGr}.
For $SO(8)$ it is amusing that 
the SW geometry with massive spinor and vector matters is 
symmetric under a part of the $SO(8)$ triality which exchanges the 
vector and spinor representations.

In sect.4, the breaking of $E_6$ to $SU(N_c)$ $(N_c \leq 6)$ is considered.
The SW geometry we will find naturally takes the form of the ALE space 
description. On the other hand, the brane dynamics relevant 
for $N=2$ $SU(N_c)$ theory
with antisymmetric matters yields the SW geometry which seems apparently
distinct from our expression \cite{LaLo,LaLoLo2}.
We will show, however, that the singularity structure exhibited by the 
complex curve in \cite{LaLo,LaLoLo2} is also realized in our 
ALE space description.

In sect.5, it is shown that the results obtained in the previous sections
can be rederived by the use of the method of $N=1$ confining phase 
superpotentials.

Finally in sect.6, we draw our conclusions.

%%%%%%%%%%%%%%%%%%%%%%%%%%%%%%%%%%%%%%%%%%%%%%%%%%%%%%%%%%%%%%%%%%%%%%
\section{Gauge symmetry breaking in Seiberg-Witten geometry}
%%%%%%%%%%%%%%%%%%%%%%%%%%%%%%%%%%%%%%%%%%%%%%%%%%%%%%%%%%%%%%%%%%%%%%

Let us consider four-dimensional $N=2$ supersymmetric gauge theory
with gauge group $G$ 
% which is simple and simply-laced, namely, $G$ is of ADE type
and $N_f$ flavors of $N=2$ hypermultiplets which consist of 
$N=1$ chiral multiplets $Q^i,\tQ_j$ ($1 \leq i,j \leq N_f$).
The $N=2$ vector multiplet contains an $N=1$ adjoint chiral multiplet $\Phi$.
Let $Q$ belong to an irreducible representation 
${\cal R}$ of the gauge group $G$ with dimension $d_R$ 
and $\tQ$ to the conjugate representation of ${\cal R}$.
The tree-level superpotential of this theory is determined
by the $N=2$ supersymmetry
\beq
W =  \sqrt{2} \sum_{i=1}^{N_f} \, \tilde{Q}_i \, \Phi_{\cal R} Q^i 
+ \sqrt{2} \sum_{i=1}^{N_f} \, m_i  \, \tilde{Q}_i Q^i,
\label{treem}
\eeq
where $\Phi_{\cal R}$ is a $d_R \times d_R$ matrix representation of $\Phi$ in 
${\cal R}$ and $m_i$ is a mass of the $i$-th hypermultiplet.

It is convenient for subsequent considerations to fix our notation 
for the root system. The simple roots of $G$ are denoted as
$\alpha_i$ where $1 \leq i \leq r$ with $r$ being the rank of $G$. Any root
is decomposed as $\alpha =\sum_{i=1}^{r} a^i \alpha_i$. The component indices
are lowered by $a_i=\sum_{j=1}^{r} A_{ij} a^j$ where $A_{ij}$ is 
the Cartan matrix. The inner product of two roots $\alpha$, $\beta$ is then
defined by
\beq
\A \cdot \B = \sum_{i=1}^{r} a^i b_i = \sum_{i,j=1}^{r} a^i A_{ij} b^j,
\eeq
where $\B=\sum_{i=1}^{r} b^i \A_i$. 

Classically the VEV of the adjoint Higgs $\Phi$ is chosen to take the 
values in the Cartan subalgebra. The classical moduli space is then 
parametrized by a Higgs VEV vector $a=\sum_{i=1}^r a^i \A_i$.
At the generic points in the classical moduli space, the gauge group $G$ is 
completely broken to $U(1)^r$. However there are singular points 
where $G$ is broken only partially to $\prod_{i} G'_i \times U(1)^l$ 
with $G'_i$ being a simple subgroup of $G$.
If we fix the gauge symmetry breaking scale to be large,
the theory becomes $N=2$ supersymmetric gauge theory with 
the gauge group $\prod_{i} G'_i \times U(1)^l$ and
the initial SW geometry reduces to the 
one describing the gauge group $G'_i$
after taking an appropriate scaling limit.

We begin with the case of $N=2$ supersymmetric
$SU(r+1)$ gauge theory with fundamental flavors.
The SW curve for this theory is given by \cite{ArFa,HaOz}
\beq
y^2={\rm det}_{r+1} \left( x-\Phi_{\cal R} \right)^2 - 
\La^{2(r+1)-N_f } \prod_{i=1}^{N_f} (m_i-x).
\eeq
Choosing the classical value $\langle \Phi_{\cal R}  \rangle_{cl}$ as
\beqa
\langle \Phi_{\cal R} \rangle_{cl} 
& =& {\rm diag} \left(\langle a^1 \rangle , \langle a^2 \ra-\la a^1 \rangle,
\langle a^3 \ra -\la a^2 \rangle, \cdots ,
\langle a^{r} \ra -\la a^{r-1} \rangle, -\langle a^{r} \rangle \right) \CR
     & = & {\rm diag} (M, M, M, \cdots , M, -r M),
\eeqa
where $M$ is a constant,
we break the gauge group $SU(r+1)$ down to $SU(r) \times U(1)$.
Note that this parametrization 
is equivalent to $\la a^j \ra = j M$ which means
$\la a_j \ra = \D_{j,r} (r+1) M$.
Setting $a_i = \D_{j,r} (r+1) M+\D a_i$ and $m_i=M+m'_i$,
we take the scaling limit $M \rightarrow \infty$ with
$\La'^{2 r-N_f}=\frac{ \La^{2(r+1)-N_f } }{(r+1) M^2}$ held fixed.
Then we are left with the SW curve corresponding to the gauge group $SU(r)$
\beq
(y')^2=\left\{ \left( x'-\D a^1 \right) 
\left( x'-(\D a^2-\D a^1) \right) \cdots
\left( x'-(-\D a^{r-1}) \right) \right\}^2-
\La'^{2 r-N_f } \prod_{i=1}^{N_f} (m'_i-x'),
% +O \left( \frac{1}{M^2} \right)
\label{sur}
\eeq
where $y'=\frac{y}{\sqrt{r+1} M}$ and $x'=x-M$.
Notice that we must shift the masses $m_i$ to obtain the finite masses 
of hypermultiplets in the $SU(r)$ theory with $N_f$ flavors.

Now we consider the case of $N=2$ theory with a simple gauge group $G$.
When we assume the nonzero VEV of the adjoint scalar, the largest non-Abelian
gauge symmetry which is left unbroken has rank $r-1$.
As we will see shortly, this largest unbroken gauge symmetry is realized 
by choosing
\beq
\la a_i \ra =M \, \D_{i,i_0}, \hskip10mm  1 \leq i \leq r,
\label{cond}
\eeq
where $M$ is an arbitrary constant and $i_0$ is some fixed value.
Under this symmetry breaking (\ref{cond}),
a gauge boson which corresponds to a generator $E_b$, where 
the subscript $b=\sum_{i} b^i \A_i$ indicates a corresponding root, 
has a mass proportional to $\la a \ra \cdot b=M \, b^{i_0}$.
This is seen from $[ \la a \ra \cdot H , E_b ] =(\la a \ra \cdot b) \, E_b$
where $H_i$ are the generators of the Cartan subalgebra.
Thus the massless gauge bosons correspond to 
the roots which satisfy $b^{i_0}=0$
and the unbroken gauge group becomes $G'_i \times U(1)$ 
where the Dynkin diagram of $G'$ is obtained by removing 
a node corresponding to the $i_0$-th simple root in the Dynkin diagram of $G$.
The Cartan subalgebra of $G$ is decomposed into 
the Cartan subalgebra of $G'$ and the additional $U(1)$ factor.
The former is generated by $E_{\A_k} \in G$ obeying
$[ E_{\A_k} , E_{\A_{-k}} ] \simeq \A_k \cdot H$ with $k \neq i_0$,
while the latter is generated by $\A_{i_0} \cdot H$. Therefore, we set
\beq
a^i = \left(A^{-1} \right)^{i \, i_0} M  + \D a^{i},
\eeq
where scalars corresponding to $G'$ have been denoted as $\D a$ 
with $\D a^{i_0} =0$. 
Note that the $U(1)$ sector decouples completely from the $G'$ sector
and the Weyl group of $G'$ naturally acts on $\D a$
out of which the Casimirs of $G'$ are constructed.

When the gauge symmetry is broken as above,
we have to decompose the matter representation ${\cal R}$ of $G$ in terms
of the subgroup $G'$ as well. We have
\beq
{\cal R} =  \bigoplus_{s=1}^{n_{\cal R}} {\cal R}_{s},
\eeq
where ${\cal R}_{s}$ stands for an irreducible representation of $G'$.
Accordingly $Q^i$ is decomposed into ${\bf Q}^i_s$
($1 \leq i \leq N_f$, $1 \leq s \leq n_{\cal R}$) 
in a $G'$ representation ${\cal R}_{s}$. $\tilde{Q}_i$ is
decomposed in a similar manner. 
After the massive components in $\Phi$ are integrated out,
the low-energy theory becomes $N=2$ $G' \times U(1)$ gauge theory.
The $U(1)$ sector decouples from the $G'$ sector and 
we consider the $G'$ sector only.
The semiclassical superpotential for this theory can be read off from 
(\ref{treem}). We have
\beq
W=\sum_{i=1}^{N_f} \left( \sqrt{2} \, \sum_{s=1}^{n_{\cal R}} 
\, ( \la a \ra \cdot \lm_{{\cal R}_{s}} +  m_i ) \, 
\tilde{{\bf Q}}_{i s}  {\bf Q}_s^i+\sqrt{2} \sum_{s=1}^{n_{\cal R}} 
\, \tilde{{\bf Q}}_{i s} \Phi_{{\cal R}_{s}} \, {\bf Q}_s^i \right) ,
\label{w1}
\eeq
where $\lm_{{\cal R}_{s}}$ is a weight of ${\cal R}$ which branches to 
the weights in ${\cal R}_{s}$.
This implies that we should shift the mass $m_i$ as 
\beq
m_i=- \la a \ra \cdot \lm_{{\cal R}_{s_i}}+m'_i=
-M {\left(\lm_{{\cal R}_{s_i}} \right) }^{i_0} +m'_i
\label{ms}
\eeq
to obtain the $G'$ theory with appropriate matter hypermultiplets.
Note that we can choose ${{\cal R}_{s_i}}$ for each hypermultiplet 
separately. This enables us to obtain the $N_f$ matters in different 
representations of $G'$ from the $N_f$ matters in a single representation 
of $G$. In the limit $M \rightarrow \infty$, 
some hypermultiplets have infinite masses and decouple from the theory.
Then the superpotential (\ref{w1}) becomes 
\beq
W=\sqrt{2} \, \sum_{i=1}^{N_f} 
\,  m'_i  \, 
\tilde{{\bf Q}}_{i \, s_i}  {\bf Q}^i_{s_i}+\sqrt{2} \sum_{i=1}^{N_f} 
\, \tilde{{\bf Q}}_{i \, s_i} \Phi_{{\cal R}_{s_i}} \, {\bf Q}^i_{s_i},
\eeq
%${\bf Q}^i={\bf Q}^i_{s_i}$ and 
%$\tilde{{\bf Q}}_{i}=\tilde{{\bf Q}}_{i \, s_i}$
and the resulting theory becomes $N=2$ theory with gauge group $G'$
with hypermultiplets belonging to the representation ${\cal R}_{s_i}$.
Note that $\la a \ra \cdot \lm_{{\cal R}_{s}}$ is proportional to 
its additional $U(1)$ charge.

In the known cases, 
the low-energy effective theory in the Coulomb phase is described
by the Seiberg-Witten geometry
which is described by a three-dimensional complex manifold 
in the form of the ALE space of ADE type fibered over ${\bf CP^1}$
\beq
z+\frac{1}{z} \La^{2 h - l({\cal R}) N_f} \prod_{i=1}^{N_f} 
{X_{G}^{{\cal R}} (x_1,x_2,x_3;a,m_i)} 
- W_G(x_1,x_2,x_3;a)=0,
\label{swg}
\eeq
where $z$ parametrizes ${\bf CP^1}$,
$h$ is the dual Coxeter number of $G$ and $l({\cal R})$
is the index of the representation  ${\cal R}$ of the matter.
Here $W_G(x_1,x_2,x_3;a)=0$ is a simple singularity of type $G$ and 
$X_{G}^{{\cal R}}(x_1,x_2,x_3;a,m_i)$ is some polynomial of 
the indicated variables. Note that
the simple singularity $W_G$ depends only on the gauge group $G$, but 
the $X_{G}^{{\cal R}}(x_1,x_2,x_3;a,m_i)$ 
depends on the matter content of the theory.

Starting with (\ref{swg}) let us consider the symmetry breaking in the 
SW geometry. In the limit $M \rightarrow \infty$, the gauge symmetry $G$
is reduced to the smaller one $G'$. The SW geometry is also reduced 
to the one with gauge symmetry $G'$ in this limit.
We can see this by substituting  $a= \la a \ra +\D a$ into (\ref{swg}) 
and keeping the leading order in $M$.
To leave the $j$-th flavor of hypermultiplets in the $G'$ theory,
its mass $m_j$ is also shifted as in (\ref{ms}).
After taking the appropriate coordinate $(x'_1,x'_2,x'_3)$ we should have 
\beqa
W_G(x_1,x_2,x_3;a) & =& 
M^{h-h'} W_{G'} (x'_1,x'_2,x'_3;\D a) + {\it o}(M^{h-h'}), \CR
X_{G}^{{\cal R}}(x_1,x_2,x_3;a, m_j) & =& 
M^{l({\cal R}) -l({\cal R}_{s_j})} X_{G'}^{{\cal R}_{s_j}} 
(x'_1,x'_2,x'_3;\D a,m'_j) 
+ {\it o}(M^{l({\cal R}) -l({\cal R}_{s_j})}), \CR
\eeqa
where $W_{G'}$ is a simple singularity of type $G'$,
$X_{G'}^{{\cal R}_{s_j}}$ is some polynomial of the indicated variables,
$h'$ is the dual Coxeter number of $G'$ and $l({\cal R}_{s_j})$ is 
the index of the representation ${\cal R}_{s_j}$ of $G'$.
The dependence on $M$ can be understood from 
the scale matching relation between theories with gauge group $G$ and $G'$
\beq
\La'^{\,\, 2 h'-\sum_{j=1}^{N_f} l({\cal R}_{s_j}) } =
\frac{ \La^{2 h- l({\cal R}) N_f } }{M^{2(h-h') - 
( l({\cal R}) N_f - \sum_{j} l({\cal R}_{s_j}) ) } },
\label{scale}
\eeq
where $\La'$ is the scale of the $G'$ theory.
Thus, in the limit $M \rightarrow \infty$, the SW geometry becomes
\beq
z'+\frac{1}{z'} \La'^{\,\, 2 h'-\sum_{j=1}^{N_f} l({\cal R}_{s_j}) }
\prod_{j=1}^{N_f} X_{G'}^{{\cal R}_{s_j}} (x'_1,x'_2,x'_3;\D a,m'_j) 
- W_{G'} (x'_1,x'_2,x'_3;\D a)=0,
\label{swg2}
\eeq
where $z'=z / M^{h-h'}$.
In the following two sections we will apply this reduction procedure explicitly
to the $N=2$ gauge theory with gauge group $E_6$ with
$N_f$ fundamental hypermultiplets.

%%%%%%%%%%%%%%%%%%%%%%%%%%%%%%%%%%%%%%%%%%%%%%%%%%%%%%%%%%%%%%%%%%%%%%
\section{Breaking $E_6$ gauge group to $SO(10)$}
%%%%%%%%%%%%%%%%%%%%%%%%%%%%%%%%%%%%%%%%%%%%%%%%%%%%%%%%%%%%%%%%%%%%%%

\begin{figure}
\epsfxsize=100mm
\hspace{3cm} \epsfbox{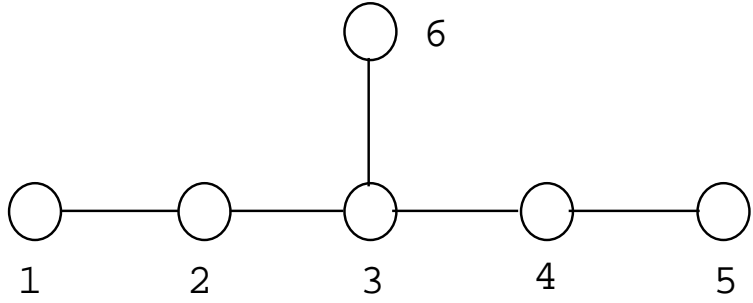}
%\vspace{-3cm}
\caption{$E_6$ Dynkin diagram}
\label{fig:E6D}
\end{figure}

There are two ways of removing a node from the Dynkin diagram of $E_6$ 
to obtain a simple group $G'$ (see fig.\ref{fig:E6D}).
When a node corresponding to $\A_5$ (or $\A_6$) is removed,
we have $G'=SO(10)$ (or $SU(6)$).
The former corresponds to the case of $G'=SO(10)$ and the latter to $G'=SU(6)$.
First we consider the breaking of $E_6$ gauge group 
down to $SO(10)$ by tuning VEV of $\Phi$ as
$\la a_i \ra = M \D_{i,5}$.
Using the inverse of the Cartan matrix we get
$\la a^i \ra = 
(\frac{2}{3} M ,\frac{4}{3} M ,\frac{6}{3} M,\frac{5}{3} M,\frac{4}{3} M,M)$.

The Seiberg-Witten geometry for $N=2$ gauge theory with 
gauge group $E_6$ with $N_f$ fundamental matters is proposed in \cite{TeYa3}
\beq
z+{1\over z} \La^{24-6N_f} 
\prod_{i=1}^{N_f} X_{E_6}^{\bf 27} (x_1,x_2,x_3;w,m_i)
- W_{E_6}(x_1,x_2,x_3;w)=0,
\label{e6ale}
\eeq
where
\beq
W_{E_6}(x_1,x_2,x_3;w)=x_1^4+x_2^3+x_3^2+w_2\, x_1^2 x_2+w_5\, x_1x_2
+w_6\, x_1^2+w_8\, x_2+w_9\, x_1+w_{12},
\eeq
and
\beqa
&& X_{E_6}^{\bf 27} (x_1,x_2,x_3;w,m_i)  \CR
&=& 8 \left( {m_i}^6+2 w_2 {m_i}^4-8{m_i}^3x_1 
+\left( w_2^2-12 x_2 \right) {m_i}^2 \right. \CR
&& \left. \hspace{3cm} 
+4 w_5 {m_i}-4 w_2 x_2 -8(x_1^2-ix_3+w_6/2) \right).
\eeqa
Here $w_k=w_k(a)$ is the degree $k$ Casimir of $E_6$ made out of $a_j$
and the degrees of $x_1,x_2$ and $x_3$ are $3,\, 4$ and $6$ respectively.
Now, substituting $a_i= M \D_{i,5}+\D a_i$ into $w_k(a)$ 
and setting $\D a^5=0$,
we expand $W_{E_6}$ and $X_{E_6}^{\bf 27}$ in $M$.
As discussed in the previous section,
there should be coordinates $( x'_1,x'_2,x'_3 )$ which can eliminate the 
terms depending upon $M^l$ $(5 \leq l \leq 12)$ in $W_{E_6}$.
Indeed, we can find such coordinates as,
\beqa
x_1 & =& -\frac{2}{27} M^3-\frac{1}{4} M x'_1 -\frac{1}{6} M w_2, \CR
x_2 & =& \frac{1}{54} M^4+\frac{1}{12} M^2 x'_1 +\frac{1}{9} M^2 w_2
+\frac{1}{8} x'_2 +\frac{1}{6} w_2^2, \CR
x_3 & =& - i \frac{1}{16} M^2 x'_3.
\label{coorda}
\eeqa
Then the $E_6$ singularity $W_{E_6}$ is written as
\beq
W_{E_6}(x_1,x_2,x_3;w)=\left( \frac{1}{4} M \right)^4 W_{D_5}(x'_1,x'_2,x'_3;v)
+{\it O}(M^3),
\label{a1}
\eeq
where 
\beq
W_{D_5}(x_1,x_2,x_3; v)
={x_1}^4+x_1{x_2}^2-{x_3}^2+v_2{x_1}^3+v_4x_1^2+v_6x_1+v_8+v_5x_2, 
\label{WD5}
\eeq
and $v_k=v_k(\D a)$ is 
the degree $k$ Casimir of $SO(10)$ constructed from $\D a_i$.
If we represent $\Phi$ as a $10 \times 10$ matrix of the
fundamental representation of $SO(10)$, we have
 $v_{2 l}=\frac{1}{2l} {\rm Tr} \Phi^{2 l}$ and $v_5=2 i {\rm Pf} \Phi$.
Thus we see in the $M \rightarrow \infty$ limit that
the SW geometry for $N=2$ pure Yang-Mills theory with 
gauge group $E_6$ becomes that with gauge group $SO(10)$.

Next we consider the effect of symmetry breaking in the matter sector.
The fundamental representation ${\bf 27}$ of $E_6$
is decomposed into the representations of $SO(10) \times U(1)$ as
\beq
{\bf 27} ={\bf 16}_{-\frac{1}{3}} \oplus {\bf 10}_{\frac{2}{3}} 
\oplus {\bf 1}_{-\frac{4}{3}},
\eeq
where the subscript denotes the $U(1)$ charge 
$ \A_5 \cdot \lm_i \, (1 \leq i \leq 27)$.
The indices of the spinor representation ${\bf 16}$
and the vector representation ${\bf 10}$ are 
four and two, respectively. Let us first take the scaling limit in such a 
way that the spinor matters of $SO(10)$ survive. Then 
the terms with $M^l$ $(l \geq 3)$ in $X_{E_6}^{27}$ must be
absent after a change of variables (\ref{coorda}) 
and the mass shift $m_i=\frac{1}{3} M + {m_{s}}_i$ (see (\ref{ms})).
In fact we find that
\beq
X_{E_6}^{\bf 27}(x_1,x_2,x_3;w,m_i) 
=M^2 X_{D_5}^{\bf 16}(x'_1,x'_2,x'_3;v,{m_s}_i)
+{\it O}(M),
\label{a2}
\eeq
where 
\beq
X_{D_5}^{\bf 16}(x_1,x_2,x_3;v,m)=m^4+\left( x_1+\frac{1}{2} v_2 \right) m^2
-m x_2+\frac{1}{2} x_3-\frac{1}{4} \left( v_4-\frac{1}{4} v_2^2 \right) 
-\frac{1}{4} v_2 x_1-\frac{1}{2} x_1^2.
\label{AD5}
\eeq
In order to make the vector matter of $SO(10)$ survive,
we shift masses as $m_i=-\frac{2}{3} M + {m_{v}}_i$. The result reads 
\beq
X_{E_6}^{\bf 27}(x_1,x_2,x_3;v,m_i) 
=M^4 X_{D_5}^{\bf 10}(x'_1,x'_2,x'_3;v,{m_v}_i)
+{\it O}(M^3),
\label{a3}
\eeq
where 
\beq
X_{D_5}^{\bf 10}(x_1,x_2,x_3;v,m)=m^2-x_1.
\eeq

Assembling (\ref{a1}), (\ref{a2}), (\ref{a3}) and
taking the limit $M \rightarrow \infty$ with 
\beq
\La^{16-4 N_s-2 N_v}_{SO(10) N_s N_v}=
2^{16+3 N_s+3 N_v} M^{-(8-2 N_s-4 N_v)} \La^{24-6 N_f}
\eeq
kept fixed, we now obtain the SW geometry for $N=2$ $SO(10)$ gauge theory with 
$N_s$ spinor and $N_v$ vector hypermultiplets
\beqa
&&z  + {1\over z} \La^{16-4 N_s-2 N_v}_{SO(10) N_s N_v} 
\prod_{i=1}^{N_s} X_{D_5}^{\bf 16}(x_1,x_2,x_3;v,{m_s}_i)
\prod_{j=1}^{N_v} X_{D_5}^{\bf 10}(x_1,x_2,x_3;v,{m_v}_j) \CR
&& \hspace{6cm} - W_{D_5}(x_1,x_2,x_3;v)=0,
\label{so10ale}
\eeqa
where $N_f=N_s+N_v$.
In the massless case $m_{s_i}=m_{v_j}=0$, our result agrees with that 
obtained from the analysis of the compactification of Type IIB string theory
on the elliptically fibered Calabi-Yau threefold \cite{AgGr}.
This is non-trivial evidence in support of (\ref{e6ale}).
Moreover the SW geometry derived in \cite{AgGr} is  only for 
the massless matters with $N_s-N_v=-2$. Here our expression is valid for 
massive matters of arbitrary number of flavors.

%%%%%%%%%%%%%%%%%%%%%%%%%%%%%%%%%%%%%%%%%%%%%%%%%%%%%%%%%%%%
\subsection{Breaking $SO(10)$ to $SO(8)$ and $SO(6)$}
%%%%%%%%%%%%%%%%%%%%%%%%%%%%%%%%%%%%%%%%%%%%%%%%%%%%%%%%%%%%

Next we examine the gauge symmetry breaking in the 
$N=2$ $SO(10)$ gauge theory with spinor matters.
When $\Phi$ acquires the VEV $\la a_i \ra = M \D_{i,1}$,
namely $\la a^i \ra = \Big(M,M,M,{M \over 2},{M \over 2}\Big)$, the gauge group
$SO(10)$ breaks to $SO(8)$. (we rename $\D a_i$ to $a_i$ henceforth.)
Note that the spinor representation of $SO(10)$ reduces to the 
spinor ${\bf 8s}$ and its conjugate ${\bf 8c}$ of $SO(8)$.
Upon taking the limit $M \rightarrow \infty$
with $a_i = \la a_i \ra +\D a_i$,
we make a change of variables in (\ref{WD5}) 
\beqa
x_1 & =& x'_1, \CR
x_2 & =& i M x'_2, \CR
x_3 & =& M x'_3.
\label{cood1}
\eeqa
In terms of these variables, the $D_5$ singularity is shown to be
\beq
W_{D_5}(x_1,x_2,x_3;v)=\left( - M^2 \right) W_{D_4}(x'_1,x'_2,x'_3;u)
+{\it O}(M),
\eeq
where 
\beq
W_{D_4}(x_1,x_2,x_3; u)
={x_1}^3+x_1{x_2}^2+{x_3}^2+u_2 {x_1}^2+v_4 x_1+u_6+2 i \tilde{v_4} x_2, 
\label{WD4}
\eeq
$u_k$ is 
the degree $k$ Casimir of $SO(8)$ constructed from $\D a_i$
and $\tilde{v_4}={\rm Pfaffian}$.
The contribution (\ref{AD5}) coming from the matters becomes
\beq
X_{D_5}^{\bf 16}(x_1,x_2,x_3;v,{m_s}_i) 
=M^2 X_{D_4}^{\bf 8s}(x'_1,x'_2,x'_3;u,{m'_s}_i)+{\it O}(M^3),
\eeq
where 
\beq
X_{D_4}^{\bf 8s}(x_1,x_2,x_3;u,m)
=m^2+\frac{1}{2} x_1-i \frac{1}{2} x_2+\frac{1}{4} u_2.
\eeq
In the above limit, we have taken ${m_s}_i=\frac{1}{2} M +{m'_s}_i$ which
corresponds to the spinor representation of $SO(8)$.
If we instead take ${m_s}_i=-\frac{1}{2} M +{m'_s}_i$, which
corresponds to the conjugate spinor representation, then 
$x_2$ is replaced with $-x_2$ in $X_{D_4}^{\bf 8s}$.

If we consider the vector matters of $SO(10)$,
we see that a change of variables (\ref{cood1}) without the shift of mass 
does not affect ${m_v}_i-x_1$.
Therefore,
in taking the limit $M \rightarrow \infty$ with 
\beq
\La^{12-2 N_s-2 N_v}_{SO(8) N_s N_v}=
M^{-(4-2 N_s)} \La^{16-4 N_s-2 N_v}_{SO(10) N_s N_v}
\eeq
being fixed, we conclude that the SW geometry for $N=2$ $SO(8)$ gauge theory
with $N_s$ spinor and $N_v$ vector flavors is
\beqa
&&z  + {1\over z} \La^{12-2 N_s-2 N_v}_{SO(8) N_s N_v}
\prod_{i=1}^{N_s} X_{D_4}^{\bf 8s}(x_1,x_2,x_3;u,{m'_s}_i)
\prod_{j=1}^{N_v} X_{D_4}^{\bf 8v}(x_1,x_2,x_3;u,{m_v}_j) \CR
&& \hspace{6cm} - W_{D_4}(x_1,x_2,x_3;u)=0,
\label{so8ale}
\eeqa
where $X_{D_4}^{\bf 8v}(x_1,x_2,x_3;u,m)=m^2-x_1$.

There is a ${\bf Z}_2$ action in the triality of $SO(8)$ which 
exchanges the vector representation and the spinor representation. Accordingly
the $SO(8)$ Casimirs are exchanged as
\beqa
v_2 & \leftrightarrow &  v_2, \CR
v_4 & \leftrightarrow &  -\frac{1}{2} v_4+3 {\rm Pf}+\frac{3}{8} v_2^2, \CR
{\rm Pf} & \leftrightarrow &   
\frac{1}{2} {\rm Pf} +\frac{1}{4} v_4-\frac{1}{16} v_2^2, \CR
v_6 & \leftrightarrow &  
v_6+\frac{1}{16} v_2^3-\frac{1}{4} v_4 v_2+\frac{1}{2} {\rm Pf} \,\, v_2.
\label{ex}
\eeqa
Thus the ${\bf Z}_2$ action is expected to
exchange $X_{D_4}^{\bf 8s}$ and $X_{D_4}^{\bf 8v}$ in (\ref{so8ale})
after an appropriate change of coordinates $x_i$.
Actually, using the new coordinates $(x'_1,x'_2)$ introduced by
\beqa
x_1 &=& -\frac{1}{2} x'_1+i \frac{1}{2}x'_2-\frac{1}{4} v_2, \CR
x_2 &=& -i \frac{3}{2} x'_1+ \frac{1}{2} x'_2-i \frac{1}{4} v_2,
\eeqa
we see that the $D_4$ singularity (\ref{WD4}) remains intact except
for (\ref{ex}) and
$X_{D_4}^{\bf 8s} \leftrightarrow X_{D_4}^{\bf 8v}$.

%Note that a ${\bf Z}_2$ action in the triality of $SO(8)$ which 
%exchange the spinor representation and the conjugate spinor representation
%is trivial.
%Thus the Seiberg-Witten geometry (\ref{so8ale}) is triality invariant.

One may further break the gauge group $SO(8)$ to $SO(6)$ following the breaking
pattern $SO(10)$ to $SO(8)$. Suitable coordinates are found 
to be $x_1=x'_1, x_2= i M x'_2$ and $x_3=M x'_3$.
The resulting SW geometry for $N=2$ $SO(6)$ gauge theory with 
$N_s$ spinor flavors and $N_v$ vector flavors is
\beqa
&&z  + {1\over z} \La^{8-N_s-2 N_v}_{SO(6) N_s N_v}
\prod_{i=1}^{N_s} (\frac{1}{2} x_2 \pm {m_s}_i)
\prod_{j=1}^{N_v} ( {m_v}_j^2-x_1 ) \CR
&& \hspace{6cm} - W_{D_3}(x_1,x_2,x_3;u)=0,
\label{so6ale}
\eeqa
where $W_{D_3}(x_1,x_2,x_3; u)
={x_1}^2+x_1{x_2}^2+{x_3}^2+u_2 {x_1}+u_4+2 i {\rm Pf} \Phi x_2$. The sign 
ambiguity in (\ref{so6ale}) arises from the two possible choices of the 
shift of masses in $SO(8)$ theory.

When $N_s=0$, it is seen that the present $SO(2N_c)$ results yield the
well-known curves for $SO(2N_c)$ theory with vector matters \cite{ArSh,Ha}.

%%%%%%%%%%%%%%%%%%%%%%%%%%%%%%%%%%%%%%%%%%%%%%%%%%%%%%%%%%%%%%%%%%%%%%
\section{Breaking $E_6$ gauge group to $SU(6)$}
%%%%%%%%%%%%%%%%%%%%%%%%%%%%%%%%%%%%%%%%%%%%%%%%%%%%%%%%%%%%%%%%%%%%%%

In this section we wish to break the $E_6$ gauge group down to $SU(6)$ 
by giving the VEV
$\la a_i \ra = M \D_{i,6}$ to $\Phi$, that is, 
$\la a^i \ra = (M,2 M,3 M,2 M,M,2 M)$.
As in the previous section,
we first substitute $a_i= M \D_{i,6}+\D a_i$ into $w_k(a)$ in (\ref{e6ale}) 
and set $\D a^6=0$.
Then we expand $W_{E_6}$ and $X_{E_6}^{\bf 27}$ in $M$, and
look for the coordinates $( x'_1,x'_2,x'_3 )$ which eliminate the 
terms depending on $M^l$ $(7 \leq l \leq 12)$ in (\ref{e6ale}).
We can find such coordinates as 
\beqa
&& x_1  = - \frac {5}{8} \,M^{2}\,{ x'_1} - 
{\displaystyle \frac {3}{4}} \,{ x'_1}\,{ w_2}, \CR
&& x_2  = {\displaystyle \frac {1}{16}} \,M^{4} 
+ ({\displaystyle \frac {1}{4}} \,{ x'_2} 
+ {\displaystyle \frac {1}{4}} \,{ x'_1}^{2} + 
{\displaystyle \frac {1}{12}} \,{ w_2})\,M^{2},
\CR
&& x_3  =\lefteqn{{\displaystyle \frac {1}{160}} \,M^{6} + ( - 
{\displaystyle \frac {1}{8}} \,{ x'_2} + {\displaystyle \frac {3
}{160}} \,{ w_2})\,M^{4}} \CR
 & & \hskip5mm + {\displaystyle \frac {1}{8}} \, ( { x'_3} - 
 \,{ x'_2}^{2} - 3 { x'_2}\,{ x'_1}^{2} - 
{ x'_2}\,{ w_2} + {\displaystyle \frac {2}{15}} \,
{ w_2}^{2} - 3 { x'_1}^{4})\,M^{2
} + {\displaystyle \frac {1}{2}} \,{ w_5}\,{ x'_1} - 
{\displaystyle \frac {1}{10}} \,{ w_6}, 
\label{ec}
\eeqa
in terms of which the $E_6$ singularity $W_{E_6}$ is represented as
\beq
W_{E_6}(x_1,x_2,x_3;w)=\left( \frac{1}{2} M \right)^6 W_{A_5}(x'_1,x'_2,x'_3;v)
+{\it O}(M^5),
\eeq
where 
\beq
W_{A_r}(x_1,x_2,x_3; v)
=x_1^r+x_2 x_3 +v_2{x_1}^{r-1}+v_3x_1^{r-2}+\cdots+v_r x_1+v_{r+1}, 
\eeq
and $v_k=v_k(\D a)$ is 
the degree $k$ Casimir of $SU(6)$ build out of $\D a_i$.
Hence it is seen in the $M \rightarrow \infty$ limit that 
the SW geometry for $N=2$ pure Yang-Mills theory with 
gauge group $E_6$ becomes that with gauge group $SU(6)$.

The fundamental representation ${\bf 27}$ of $E_6$
is decomposed into the representations of $SU(6) \times U(1)$ as
\beq
{\bf 27} ={\bf 15}_{0} \oplus {\bf 6}_{1} \oplus {\bf \bar{6}}_{-1},
\eeq
where the subscript denotes the $U(1)$ charge 
$ \A_6 \cdot \lm_i \, (1 \leq i \leq 27)$.
The indices of the antisymmetric representation ${\bf 15}$
and the fundamental representation ${\bf 6}$ are 
four and one, respectively.
Thus the terms with $M^l$ $(l \geq 3)$ in $X_{E_6}^{\bf 27}$ must be
absent after taking the coordinates $( x'_1,x'_2,x'_3 )$ defined in (\ref{ec}).
Note that there is no need to shift the mass
to make the antisymmetric matter survive. 
We indeed obtain a desired expression
\beq
X_{E_6}^{\bf 27}(x_1,x_2,x_3;w,m_i) 
=-M^2 X_{A_5}^{\bf 15}(x'_1,x'_2,x'_3;v,m_i)+{\it O}(M),
\eeq
where 
\beqa
&&X_{A_5}^{\bf 15}(x_1,x_2,x_3;v,m)  = 
m^4-2 m^3 x_1+ 3 \left( \frac{1}{3} v_2 +x_1^2+x_2 \right) m^2
\CR && \hspace{3cm}
+m v_3-x_3+x_1^4+2 v_2 x_1^2+3 x_2 x_1^2+v_3 x_1+x_2^2+v_2 x_2+v_4.
\eeqa
If we shift the mass as $m_i=M + {m_{f}}_i$
in order to make the vector matter survive,
we find that
\beq
X_{E_6}^{\bf 27}(x_1,x_2,x_3;v,m_i) 
=2 M^5 X_{A_5}^{\bf 6}(x'_1,x'_2,x'_3;v,{m_f}_i)+{\it O}(M^4),
\label{aa}
\eeq
where $X_{A_5}^{\bf 6}(x_1,x_2,x_3;v,m)=m+x_1$.
The shift of masses $m_i=-M + {m_{f}}_i$ also
corresponds to making the vector matter survive, but
the factor $(-1)$ is needed in the RHS of (\ref{aa}).

{}From these observations we can obtain the SW geometry for $N=2$ $SU(6)$ 
gauge theory with $N_a$ antisymmetric and $N'_f$ fundamental matters
by taking the limit $M \rightarrow \infty$ while
\beq
\La^{12-4 N_a-N'_f}_{SU(6) N_a N'_f}=
(-1)^{N_a} 2^{12+2 N'_f} M^{-(12-2 N_a-5 N'_f)} \La^{24-6 N_f}
\eeq
held fixed. Our result reads
\beqa
&&z  + {1\over z} \La^{12-4 N_a-N'_f}_{SU(6) N_a N'_f} 
\prod_{i=1}^{N_a} X_{A_5}^{\bf 15}(x_1,x_2,x_3;v,{m_a}_i)
\prod_{j=1}^{N'_f} X_{A_5}^{\bf 6}(x_1,x_2,x_3;v,{m_f}_j) \CR
&& \hspace{6cm} - W_{A_5}(x_1,x_2,x_3;v)=0,
\label{su6ale}
\eeqa
where $N_f=N_a+N'_f$.

%%%%%%%%%%%%%%%%%%%%%%%%%%%%%%%%%%%%%%%%%%%%%%%%%%%%%%%%%%%%
\subsection{Breaking $SU(6)$ to $SU(5)$, $SU(4)$ and $SU(3)$}
%%%%%%%%%%%%%%%%%%%%%%%%%%%%%%%%%%%%%%%%%%%%%%%%%%%%%%%%%%%%

We are now able to break $SU(r+1)$ gauge group to $SU(r)$ successively
by putting $\la a_i \ra = M \D_{i,r}$.
In sect.2 we have seen that 
the proper coordinates are chosen to be $x_1=x'_1+M/(r+1), x_2=x'_2$ and 
$x_3=M x'_3$ in terms of which  
$W_{A_{r}}(x_1,x_2,x_3; v)=M W_{A_{r-1}}(x'_1,x'_2,x'_3; v')+{\it O}(M^0)$.
Note that the degrees of $x_1, x_2$ and $x_3$ are $1,2$ and $r-1$, 
respectively. The antisymmetric representation of $SU(r+1)$
is decomposed into the antisymmetric and fundamental representations of 
$SU(r) \times U(1) $ as follows
\beq
{\bf \frac{r (r+1)}{2} } =
{\bf \frac{(r-1) r}{2}}_{\frac{2}{r+1}} \oplus {\bf r}_{-\frac{r-1}{r+1}},
\eeq
where the subscript denotes the $U(1)$ charge.
After some computations we can see that
the SW geometry for $N=2$ $SU(r+1)$ $(r \leq 5)$ gauge theory 
with $N_a$ antisymmetric and $N'_f$ fundamental hypermultiplets
turns out to be
\beqa
&&z  + {1\over z} \La^{2(r+1)-(r-1) N_a-N'_f}_{SU(r+1) N_a N'_f} 
\prod_{i=1}^{N_a} X_{A_r}^{\bf \frac{r(r+1)}{2}}(x_1,x_2,x_3;v,{m_a}_i)
\prod_{j=1}^{N'_f} (x_1-{m_f}_j) \CR
&& \hspace{6cm} - W_{A_r}(x_1,x_2,x_3;v)=0,
\label{sur+1ale}
\eeqa
where $X_{A_r}^{\bf \frac{r(r+1)}{2}}$ is defined as
\beq
X_{A_r}^{\bf \frac{r(r+1)}{2}} \left( x_j;v,{m_a}_i=\frac{2 M}{r+1}
+{m'_a}_i \right)
= M X_{A_{r-1}}^{\bf \frac{(r-1)r}{2}}(x'_j;v',{m'_a}_i) +{\it O}(M^0),
\eeq
and $\La^{2(r+1)-(r-1) N_a-N'_f}_{SU(r+1) N_a N'_f}
=M^{2-N_a} \La^{2 r-(r-2) N_a-N'_f}_{SU(r) N_a N'_f}$.
Explicit calculations yield
\beqa
X_{A_4}^{\bf 10}(x_j;v,{m_a}_i) \!\!\!& =&\!\!\! m^3-m^2 x_1
+(2 x_2+2 x_1^2+v_2) m
+2 x_1^2-x_3+x_2 x_1  +v_2 x_1+v_3, \CR
X_{A_3}^{\bf 6}(x_j;v,{m_a}_i) \!\!\!& =&\!\!\! m^2+x_2-x_3+2 x_1^2+v_2, \CR
X_{A_2}^{\bf 3}(x_j;v,{m_a}_i) \!\!\!& =&\!\!\! m+x_1-x_3. 
\eeqa
We also see that 
\beqa
&& X_{A_5}^{\bf 15}\Big(x_j;v,{m_a}_i=-\frac{2}{3} M+{m'_f}_i \Big)
=M^3 (x'_1-{m'_f}_i)
+{\it O}(M^2), \CR
&& X_{A_4}^{\bf 10}\Big(x_j;v,{m_a}_i=-\frac{3}{5} M+{m'_f}_i \Big)
=-M^2 (x'_1-{m'_f}_i)
+{\it O}(M^1), \CR
&& X_{A_3}^{\bf 6}\Big(x_j;v,{m_a}_i=-\frac{1}{2} M+{m'_f}_i \Big)
=M (x'_1-{m'_f}_i-x'_3)+{\it O}(M^0)
\eeqa
by shifting masses in such a way that the fundamental matters remain.

We now check our $SU(N_c)$ results. First of all, 
for $SU(3)$ gauge group, the antisymmetric representation is identical to
the fundamental representation. 
Thus (\ref{sur+1ale}) should be equivalent to the well-known $SU(3)$ curve.
In fact, if we integrate out variables $x_2$ and $x_3$, 
the SW geometry (\ref{sur+1ale}) yields the $SU(3)$ curve with
$N_a+N_f'$ fundamental flavors.

Let us next turn to the case of $SU(4)$ gauge group.
Since the Lie algebra of $SU(4)$ is the same as that of $SO(6)$,
the antisymmetric and fundamental representations of $SU(4)$ correspond to 
the vector and spinor representations of $SO(6)$ respectively.
This relation is realized in (\ref{sur+1ale}) and (\ref{so6ale}) as follows.
If we set $x_1=\frac{1}{2} x'_2$, $x_2=i x'_3- \frac{1}{2}x'_1-
\frac{1}{4} {x'_2}^2-\frac{1}{2} v_2$ and $x_3=i x'_3+\frac{1}{2} x'_1+
\frac{1}{4} {x'_2}^2+\frac{1}{2} v_2$,
we find 
\beq
W_{A_3}(x_i;v)=-\frac{1}{4} W_{D_3} (x'_i;u),
\eeq
where $u$ is related to $v$ through
$u_2=2 v_2, u_4=-4 v_4+ v_2^2$ and ${\rm Pf} =i v_3$.
Moreover we obtain $X_{A_3}^{\bf 6}(x_j;v,{m_a}_i) = {m_a}_i^2-x'_1$ and
$x_1-{m_f}_j=\frac{1}{2} x'_2-{m_f}_j$. Thus our $SU(4)$ result is in
accordance with what we have anticipated. This observation provides a
consistency check of our procedure since both $SO(6)$ and $SU(4)$ results
are deduced from the $E_6$ theory via two independent routes associated
with different symmetry breaking patterns.

Checking the $SU(5)$ gauge theory result is most intricate. 
Complex curves describing $N=2$ $SU(N_c)$ gauge theory with
matters in one antisymmetric representation and fundamental representations
are obtained in \cite{LaLo,LaLoLo2} using brane configurations.
Let us concentrate on $SU(5)$ theory with one massless antisymmetric matter
and no fundamental matters in order to compare with our result (\ref{so6ale}).
The relevant curve is given by \cite{LaLo}
\beqa
& & y^3+x y^2 (x^5+v_2 x^3-v_3 x^2+v_4 x-v_5) \CR
& & \hspace{1.5cm} 
+y \La^7 (3 x^5+3 v_2 x^3-v_3 x^2+3 v_4 x-v_5) +2 \La^{14} (x^4+v_2 x^2+v_4)=0.
\label{curL}
\eeqa
The discriminant of (\ref{curL}) has the form
\beq
\Delta_{Brane}
=F_0(v) \La^{105} (27 \La^7v_4^2+v_5^3) ( H_{50}(v,L) )^2 ( H_{35}(v,L) )^6,
\eeq
where $F_0$ is some polynomial in $v$, $H_{n}$ is a degree $n$ 
polynomial in $v$ and $L=-\La^7/4$.
If we set $v_2=v_3=0$ for simplicity, then
\beqa
H_{50}(v,L) &=& 65536\,{v_4}^{10}\,{v_5}^{2} + 1048576\,{v_4}^{
9}\,{L}^{2} - 33587200\,{v_4}^{7}\,{v_5}^{3}\,{L} 
+ 1600000\,
{v_4}^{5}\,{v_5}^{6} \CR
& & - 539492352\,{v_4}^{6}\,{v_5}\,{L}^{3} + 
3261440000\,{v_4}^{4}\,{v_5}^{4}\,{L}^{2} + 390000000\,
{v_4}^{2}\,{v_5}^{7}\,{L} \CR 
& & + 9765625\,{v_5}^{10} + 143947517952\,{v_4}^{3}\,{v_5}^{2}\,{L}^{4} + 
5378240000\,{v_4}\,{v_5}^{5}\,{L}^{3} \CR
& & + 1457236279296\, {v_4}^{2}\,{L}^{6} + 53971714048\,{v_5}^{3}\,{L}^{5}, \CR
H_{35}(v,L)& =& 32 v_5^7+432 L v_4^2 v_5^2+17496 L^3 v_4 v_5^2+177147 L^5.
\eeqa
We have also calculated the discriminant $\Delta_{ALE}$ of our expression
(\ref{sur+1ale}) with $r=4$ and found it in the factorized form.
Evaluating $\Delta_{Brane}$ and $\Delta_{ALE}$ at sufficiently many points 
in the moduli space, we observe that $\Delta_{ALE}$ also contains a factor
$H_{50}(v,L)$ with $\La_{SU(4) 1,0}^7=L$. This fact may be regarded as a
non-trivial check for the compatibility of the M-theory/brane dynamics 
result and our ALE space description. It is thus inferred that only the zeroes
of a common factor $H_{50}(v,L)$ in the discriminants represent the
physical singularities in the moduli space.\footnote{A similar phenomenon
is observed in $SU(4)$ gauge theory. We have checked that the discriminant 
of the curve for $SU(4)$ theory with one massive antisymmetric hypermultiplet
proposed in \cite{LaLo} and that of our ALE formula (\ref{sur+1ale}) 
with $r=3$ carry a common factor.}

%%%%%%%%%%%%%%%%%%%%%%%%%%%%%%%%%%%%%%%%%%%%%%%%%%%%%%%%%%%%%%%%%%%%%%
\section{N=1 Confining phase superpotentials}
%%%%%%%%%%%%%%%%%%%%%%%%%%%%%%%%%%%%%%%%%%%%%%%%%%%%%%%%%%%%%%%%%%%%%%

In this section we will rederive the SW geometry 
obtained in the previous sections using the 
method of $N=1$ confining phase superpotentials.
We will explain the essence of this method in the following.
More detailed explanation 
is presented in \cite{ElFoGiInRa,TeYa2,TeYa3}.
First we add a tree-level superpotential 
\beq
W_{tree}=\sum_{i=1}^r g_i s_i(\Phi),
\eeq
to perturb $N=2$ theory with gauge group $G$ to $N=1$ theory,
where $s_i(\Phi)$ are Casimirs of $G$ built out of $\Phi$ and
$g_i$ are coupling parameters. It is then observed that only the 
singularities of the moduli space where dyons become
massless remain as the $N=1$ vacua.
Thus studying this perturbed $N=1$ theory with a confined photon,
which corresponds to unbroken $SU(2) \times U(1)^{r-1}$ vacua classically, 
we can find the physical singular loci of $N=2$ moduli space and 
construct the corresponding SW geometry.
Recall that a basis of Casimirs $s_i$ should be chosen judiciously to 
obtain the correct results.
The SW geometry (\ref{e6ale}) is derived in this manner \cite{TeYa3}.

Following \cite{TeYa3} let us now describe the computation in the $N=1$
confining phase approach. In all the cases considered below, we use $w_i$ 
to denote the deformation parameters of the standard $ADE$ singularities. 
(In the previous sections, we have used $u_i$ for $SO(2 r)$ and 
$v_i$ for $SU(r+1)$ instead of $w_i$.)
$P (y_i, s_i)$ is defined as a polynomial which becomes zero 
if we evaluate $s_i$ and $y_i$ in 
the classical $SU(2) \times U(1)^{r-1}$ vacua and
$X (y_i, s_i, m)$ stands for the ``matter factor'' which relates 
the scale of the high-energy theory to that of the low-energy $SU(2)$ 
Yang-Mills theory taking into account the factor arising from the Higgs 
effect \cite{TeYa2}. The SW geometry is obtained as \cite{TeYa3}
\beq
z+\frac{1}{z} \La^{2 h - l({\cal R}) N_f} X^{N_{f}} + P=0,
\eeq
where $l({\cal R})$ is the index of the representation ${\cal R}$ of the matter
and $N_{f}$ denotes the number of matters in ${\cal R}$.

\subsection{SU(r+1) gauge theory}

In this subsection it is convenient to put $y_n=g_{r-n}/g_r$.
In $SU(4)$ theory, we should take 
\beq
s_2 = w_2, \;\; s_3=w_3, \;\; s_4=w_4-\frac{1}{4} {w_2}^2.
\eeq
Then we obtain
${ P} =  - { y_2}^{2} + 2\,{ y_2}\,{ y_1}^{2} 
+ { s_2}\,{ y_2} + { s_3}\,{ y_1} + { s_4}$.
For hypermultiplets in the antisymmetric representation, we get 
${ X} = m^{2} + 2 y_2$, thereby the ALE expression of the SW geometry
is reproduced. On the other hand, 
an $N=1$ confining phase superpotential is considered in \cite{OdToSaSa}
so as to produce the curve based on an M-fivebrane configuration \cite{LaLo}.

In $SU(5)$ theory, we should take 
\beq
s_2=w_2, \;\; s_3=w_3, \;\; s_4=w_4-\frac{1}{4} {w_2}^2,
 \;\; s_5=w_5-\frac{1}{2} {w_2} {w_3}.
\eeq
Then we obtain  ${ P} = 
2\,{ y_1}^{2}{ y_3}  - 2\,{ y_2} { y_3}+{ y_2}^{2}{ y_1}
+ { s_2}\,{ y_3} + { s_3}\,{ y_2} + 
{ s_4} { y_1} +{ s_5}$.
Considering the antisymmetric flavors,
we get ${ X} = m^{3} -{y_1} m^2+2 m {y_2}+ 2 y_3$.

In $SU(6)$ theory, we should take 
\beqa
&&s_2=w_2, \;\; s_3=w_3, \;\; s_4=w_4-\frac{1}{4} {w_2}^2,  \;\;  \CR
&&s_5=w_5-\frac{1}{2} {w_2} {w_3}, 
\;\; s_6=w_6-\frac{1}{2} {w_2} {w_4} % -\frac{1}{4} {w_3}^2
+\frac{1}{8} {w_2}^3.
\eeqa
In this case $y_3=y_2 y_1$ and we find ${ P} = 
2 { y_1}^{2} { y_4} -2 {y_2}{y_4}-{y_2}^2 {y_1}^2
+ { s_2}\,{ y_4} + { s_3}\,{ y_2} { y_1}+ 
{ s_4} { y_2} +{ s_5} { y_1} +{ s_6}$.
For the antisymmetric flavors,
we obtain ${ X} =m^{4} - 2\,{ y_1}\,m^{3} 
+ m^{2}\,({ y_1}^2+2 { y_2} )+ 
+ m\,s_3 + { y_2}^2+2\,{ y_4}$.

%and for the (00100) representation (the dimension is 20 and index is 6),
%\beq
%{ Xb} := m^{6} + m^{4}\,(4\,{ y_2} - { y_1}^{2}) + m^{2}
%\,( - 8\,{ y_4} - 4\,{ y_3}\,{ y_1}) - 4\,{ y_3}^{2}.
%\eeq

In all the cases above, we see that the SW geometry obtained from 
the confining phase superpotential method
is in agreement with (\ref{sur+1ale}) after an appropriate change of
coordinates.

\subsection{SO(2r) gauge theory}

In $SO(8)$ theory, we should take 
\beqa
&&s_2=w_2, \;\; s_4=w_4-\frac{1}{3} {w_2}^2,  \;\;  
s_6=w_6-\frac{1}{3} {w_4} {w_2} +\frac{2}{27} {w_2}^3, \CR
&&s'_4={\rm Pfaffian}.
\eeqa
Let  ${y_1}=g'_4/g_6$, $y_2=g_4/g_6$ and $y_3=g_2/g_6$.
In this case ${y_3}=\frac{1}{12} {y_1}^3$ and we find 
${ P} ={y_2}^3-\frac{1}{4} {y_1}^2 {y_2} +\frac{1}{12} s_2 {y_1}^2+
s'_4 {y_1} +s_4 y_2 +s_6$.
If we consider the antisymmetric flavors,
we get 
\beq
{ X} =m^2+\frac{1}{4} {y_1}+\frac{1}{2} {y_2} +\frac{1}{12} s_2,
\eeq
and for the fundamental representation 
\beq
{ X_f} = m^2-{y_2} +\frac{1}{3} s_2.
\eeq
Using the relation $y_1=-2 i x_2$ and $y_2=x_1+\frac{1}{3} v_2$ 
one can see that this result is equivalent to (\ref{so8ale}).

In $SO(10)$ theory, 
we should take 
\beqa
&&s_2=w_2, \;\; s_4=w_4-\frac{1}{4} {w_2}^2,  \;\;  
{\rm s_5}={\rm Pfaffian}, \;\;
s_6=w_6-\frac{1}{2} {w_4} {w_2} +\frac{1}{8} {w_2}^3, \CR
&&s_8=w_8-\frac{1}{4} {w_4}^2 +\frac{1}{8} {w_4} {w_2}^2- \frac{1}{64} {w_2}^4.
\eeqa
Let ${y_2}=g_5/g_8$, 
${y_1}=g_6/g_8$, ${y_3}=g_4/g_8$ and ${y_4}=g_2/g_8$.
There is a relation ${y_4}={y_1} {y_3}$ and we find $
{ P} =-{y_3}^2+2 y_3 {y_1}^2 + \frac{1}{4} {y_2}^2 {y_1}
+ s_2 {y_3} {y_1}+s_4 y_3+ s_5 y_2 + s_6 y_1 +s_8$.
If we consider the antisymmetric flavors, we get 
${ X} =m^4+m^2 (\frac{1}{2} s_2+ y_1)-m \frac{1}{2} {y_2} -\frac{1}{2} y_3$.
For the fundamental representation, ${ X_f} = m^2-{y_1}$.
Similarly to the case of $SO(8)$, we can verify the 
equivalence of this result to (\ref{so10ale}).

%%%%%%%%%%%%%%%%%%%%%%
\section{Conclusions}
%%%%%%%%%%%%%%%%%%%%%%

Starting with the SW geometry for $N=2$ supersymmetric gauge theory with
gauge group $E_6$ with massive fundamental hypermultiplets, we have obtained 
the SW geometry for $SO(2 N_c)$ $(N_c \leq 5)$ theory with massive spinor
and vector hypermultiplets by implementing the gauge symmetry breaking in
the $E_6$ theory. The other symmetry breaking pattern has been used to
derive the SW geometry for $N=2$ $SU(N_c)$ $(N_c \leq 6)$ theory with
massive antisymmetric and fundamental hypermultiplets. All the SW geometries
we have obtained are of the form of ALE fibrations over a sphere.
Whenever possible our results have been compared with those obtained in
the approaches based on the geometric engineering and the brane 
dynamics. It is impressive to find an agreement in spite of the fact that
the methods are fairly different. Furthermore the SW geometries derived from 
the $E_6$ theory have also been obtained from the point of view 
of $N=1$ confining phase superpotentials.

Let us mention here that the SW geometry for $N=2$ $E_7$ gauge theory
with massive fundamental matters will be obtained without any essential
difficulty by finding an appropriate $N=1$ confining phase superpotential.
The symmetry breaking of $E_7$ will then give rise to $SO(12)$ theory with 
spinor matters as well as $SU(7)$ theory with antisymmetric matters.
For the gauge group $E_8$, however, the situation seems rather subtle in
employing the confining phase superpotential technique since there is no
distinction between the fundamental and adjoint representations.

Finally, in order to analyze the mass of the BPS states and 
other interesting properties of 
the theory, one has to know the Seiberg-Witten three-form and 
appropriate cycles in the ALE fibration space.
For $N=2$ $SO(10)$ theory with massless spinor and vector hypermultiplets,
these objects may be obtained in principle from the 
Calabi-Yau threefold on which the string theory is compactified \cite{AgGr}.
It is important to find the SW three-form and appropriate cycles for the
SW geometry when the massive hypermultiplets exist. This issue is
left for our future consideration.

\vskip4mm\noindent
{\bf Acknowledgements}

\vskip2mm
The work of S.T. is supported by JSPS Research Fellowship for Young 
Scientists. The work of S.K.Y. was supported in part by Grant-in-Aid 
for Scientific Research from the Ministry of Education, Science and Culture
(No. 09640335).

\newpage

%%%%%%%  References

\end{document}